\documentstyle[sprocl]{article}
\begin{document}
\def\be{\begin{equation}}
\def\ee{\end{equation}}
\def\bea{\begin{eqnarray}}
\def\eea{\end{eqnarray}}
\def\simlt{\stackrel{<}{{}_\sim}}
\def\simgt{\stackrel{>}{{}_\sim}}
\def\Im{\mathop{\rm Im}}
\def\ov{\overline}
\def\npb#1#2#3{{\it Nucl.~Phys.} {\bf{B#1}} (19#3) #2}
\def\plb#1#2#3{{\it Phys.~Lett.} {\bf{B#1}} (19#3) #2}
\def\prd#1#2#3{{\it Phys.~Rev.} {\bf{D#1}} (19#3) #2}
\def\prl#1#2#3{{\it Phys.~Rev.~Lett.} {\bf{#1}} (19#3) #2}
\def\ZPC#1#2#3{{\it Z.~Phys.} {\bf C#1} (19#2) #3}
\def\PTP#1#2#3{{\it Prog.~Theor.~Phys.} {\bf#1}  (19#2) #3}
\def\MPLA#1#2#3{{\it Mod.~Phys.~Lett.} {\bf#1} (19#2) #3}
\def\PR#1#2#3{{\it Phys.~Rep.} {\bf#1} (19#2) #3}
\def\AP#1#2#3{{\it Ann.~Phys.} {\bf#1} (19#2) #3}
\def\RMP#1#2#3{{\it Rev.~Mod.~Phys.} {\bf#1} (19#2) #3}
\def\HPA#1#2#3{{\it Helv.~Phys.~Acta} {\bf#1} (19#2) #3}
\def\JETPL#1#2#3{{\it JETP~Lett.} {\bf#1} (19#2) #3}
\def\Re{\mathop{\rm Re}}
\def\Tr{\mathop{\rm Tr}}
\def\und{\underline}
\def\dalpha{{\dot\alpha}}
\def\dbeta{{\dot\beta}}
\def\drho{{\dot\rho}}
\def\dsigma{{\dot\sigma}}
\def\crbig{\\\noalign{\vspace {3mm}}}
\def\bigint{{\displaystyle\int}} 
\def\Fcomp{{\theta\theta}}
\def\Fbarcomp{\ov{\theta\theta}}
\def\Dcomp{{\theta\theta\ov{\theta\theta}}}
\def\Dint{{\bigint d^2\theta d^2\ov\theta\,}}
\def\Fint{{\bigint d^2\theta\,}}
\def\Fbarint{{\bigint d^2\ov\theta\,}}
\def\ex{{\rm exp}}
\def\mst11{m_{\;\widetilde{t}_{1}}}
\def\msti{m_{\;\widetilde{t}_i}}
\def\mstj{m_{\;\widetilde{t}_j}}
\def\msbi{m_{\;\widetilde{b}_i}}
\def\msbj{m_{\;\widetilde{b}_j}}
\def\st{\;\widetilde{t}}
\def\sb{\;\widetilde{b}}
\def\mst22{m_{\;\widetilde{t}_{2}}}
\def\mst12{m_{\;\widetilde{t}_{1,2}}}
\def\mstl{m_{\;\widetilde{t}_L}}
\def\mstr{m_{\;\widetilde{t}_R}}
\def\msb11{m_{\;\widetilde{b}_{1}}}
\def\msb22{m_{\;\widetilde{b}_{2}}}
\def\msb12{m_{\;\widetilde{b}_{1,2}}}
\def\msbl{m_{\;\widetilde{b}_L}}
\def\msbr{m_{\;\widetilde{b}_R}}
\def\modh{\left|H\right|}
\def\mwidetilde2{\widetilde{m}^{2}}
\def\lambdawidetilde{\widetilde{\lambda}}
\def\Lambdawidetilde{\widetilde{\Lambda}}
\def\leff{\lambda_{\rm eff}}
\def\mbart{\overline{m}_{t}}
\def\exis{\varphi}
%%%%%%%%%%%%%%%%%%%%%%%%%%%%%%%%%%%%%%%%%%%%%%%
%%%%%%%%%%%%%%%%%%%%%%%%
%\def\baselinestretch{1}
%
% \hoffset = .65 in\begin{document}
 \pagestyle{empty}
\begin{flushright}
 MIT-CTP-2699 \\
hep-ph/9711471
\end{flushright}
\vspace*{5mm}
\begin{center}
{\Large \bf  Supersymmetry and Inflation ~\footnote{To appear in {\it Perspectives on
Higgs Physics II}, ed. G.L. Kane, World Scientific, Singapore}}\\
\vspace{1.5cm}
{\large L. Randall }  \\
\vspace{0.3cm}
   Massachusetts Institute of Technology  \\
 Center for Theoretical Physics \\
77 Massachusetts Avenue\\
Cambridge, MA 02139 \\
 \vspace*{2cm}
Abstract
\end{center}
Inflation is a promising solution to many problems of the standard Big-Bang
cosmology. Nevertheless, inflationary models have proved less compelling.
In this chapter, we discuss why supersymmetry  has led to more  natural
models of inflation.  We pay particular attention to multifield models,
both with a high and a low Hubble parameter.
 \vfill
\begin{flushleft}
 November 1997
\end{flushleft}
\eject
\pagestyle{empty}
%\clearpage\mbox{}\clearpage
\setcounter{page}{1}
\pagestyle{plain}
\newpage
\title{ Supersymmetry and Inflation}
\author{ L. Randall},
\address{ Massachusetts Institute of Technology, Center for
Theoretical Physics, 77 Massachusetts Avenue,
Cambridge, MA 02139}
  \maketitle\abstracts{
Inflation is a promising solution to many problems of the standard Big-Bang
cosmology. Nevertheless, inflationary models have proved less compelling.
In this chapter, we discuss why supersymmetry  has led to more  natural
models of inflation.  We pay particular attention to multifield models,
both with a high and a low Hubble parameter.
}

\section{Introduction}
   Supersymmetric cosmology is necessarily a speculative subject,
since the evolution of the universe is sensitive    not only to
the observed light degrees of freedom and their superpartners,
but also to the as yet undetected heavy particle spectrum.
  The heavy degrees of freedom
only  decouple at low  temperatures; in the early universe
they can be very relevant.  In fact,
degrees of freedom which are heavy could
have been light in the early universe, and vice versa.
Furthermore, the vacuum structure
of supersymmetric and superstring  theories can be very rich and complex;
we do not yet know how  our vacuum is determined. 
Nevertheless, despite our ignorance of many aspects of
high-energy particle physics, there are certain features
of supersymmetric theories which have been shown in
recent years to be relevant to cosmology.    If the
world is supersymmetric, it  is clearly  important
for cosmology, both because of the many new particles which
would be present
and because of the  many flat direction (moduli) fields.  
These are fields which have no potential in the supersymmetric
limit.  They can however get a small potential due
to supersymmetry breaking, higher dimension operators,
or interactions with other fields.
These flat directions which only occur naturally in supersymmetric
theories can provide large amounts of energy as they
will almost certainly not start their evolution from
their minimum. Many recent models of inflation are based
on this observation, though often in different contexts.
 In this chapter, we  will see how supersymmetric theories might
provide more compelling models of inflation. We will consider
some   examples which demonstrate
that supersymmetric theories might provide viable inflaton
candidates. Even without knowing the correct particle physics
model at high energy, we can identify what might be desirable features
of this model if they are to simultaneously account for an
earlier epoch of inflation.

In this chapter we will briefly review the motivation for
inflation and the requirements for a successful inflationary
cosmology \cite{turner,linde}. We will then discuss in some detail the multifield
models of inflation which potentially succeed in meeting the requirements
of inflation with little or no fine-tuning. We will discuss
several particular models, but cannot attempt a complete
enumeration of all models to date-this list is changing
very rapidly! We will instead focus on what we think
are the additional requirements of the supersymmetric
inflation models, their possible predictions, and important
questions which remain and attempts to address them.

The standard Big Bang cosmology has many important successes.
Most notable are the measured Hubble expansion of the universe,
the predictions of the light element abundances from nucleosynthesis
in the early universe, and the prediction of the $2.7^\circ$
microwave background radiation spectrum.  More
recently, the measurment of the anisotropy in the cosmic
microwave background is an indication that
theories of structure formation are on the right track. The standard
cosmology is simple and successful, but very likely incomplete.
As with the standard model
of  particle physics, the major reason this is believed is
that the model as it stands is unnatural, in that it
requires very fine-tuned initial conditions.

The shortcomings of the standard cosmology are the
problems of the large-scale smoothness of the universe,
the spatial flatness   problems, the origins
of small inhomogeneities, and the potential presence
of unwanted relics.  Inflationary cosmology successfully
resolves the first two problems for a  sufficiently long-lived
inflationary phase. If inflation involves the correct
mass scales and/or parameters, inflation can also
lead to the observed density perturbations. For relics
which   do not get produced
late in the universe, inflation can solve the problem
of unwanted relics, although it should be noted that
the problem of unwanted relics is  a serious consideration
for most supersymmetric theories, even with an early
inflationary epoch.

Most successful inflationary models are based on slow-roll
inflation \cite{as,linde1}.  In its earliest implementation it is phenomenologically
successful as a model of inflation but requires fine-tuning,
either of the potential or of the initial conditions.\footnote{There
is debate over whether an initial condition should be considered
fine-tuned, particularly in an eternal inflationary scenario.
We will not discuss this here but refer the reader to Ref. \cite{linde}.}
The requirements on the potential for the inflaton field $\phi$
for slow roll to be valid are $|V''(\phi)|<9H^2$ and $|V'M_P/V| <\sqrt{48 \pi}$.
While these constraints are met, the potential is approximately
constant as is the Hubble parameter, $H=\sqrt{8 \pi V/3M_P^2}$.
During the period of slow-roll, the universe can expand by
an exponential factor, the value of which is determined
by the time for which the slow-roll conditions are valid.  Inflation
ends when these conditions cease to apply.

So far, we see that the important condition for inflation is
to have an approximately constant energy density for a finite
interval.  This in and of itself is not a serious constraint,
particularly in a supersymmetric theory for which many
light or massless scalars might be present.  What makes
the construction of inflationary models tricky (or fine-tuned)
is that inflation needs to end,
so that  reheating can produce the known matter content of the universe.
With the further requirement
that inflation accounts for the density fluctuations in the microwave background (here given in the slow-roll approximation), $V^{3/2}/V'M_P^3=5.4 \cdot 10^{-4}$, 
one is led to the introduction of small parameters. These
problems have been reviewed elsewhere \cite{turner,linde}.  
In general,
\begin{equation}
{\delta \rho \over \rho} \sim {\delta  N}  \sim {H^2 \over \dot{\phi}}
\sim {H^3 \over V'}\sim {H^3 \over m^2 \phi}
\end{equation}
where use has been made of the slow-roll equation of
motion $3H \dot{\phi}=-V'(\phi)$ and $N$ is the number
of e-folds.  
 It is important
to notice that the Hubble parameter which will give the
correct magnitude of density perturbations is determined
not only by $m$, the mass of the inflaton, but also by $\phi$,
which in this case means the magnitude of the field $\phi$
during the time density fluctuations are formed.  Therefore,
although it is conventionally assumed that $H$ during inflation
is determined to get the density fluctuations right,
this is not necessarily the case. By constructing an inflationary
model with a different value of $\phi$, one can obtain
more than one scale for $H$ which can yield sensible density perturbations.

We note that it is not an essential requirement that density fluctuations
formed during inflation account for the observed structure. Other
suggestions for producing density perturbations have been given \cite{brand}.
However, it would certainly be more economical and greatly desirable
to have density perturbations taken care of during inflation, since
inflation automatically produces fluctuations. This
is the assumption which we make here. Moreover, recent
papers indicate a substantial vector and tensor contribution to the CMBR
implying too low anisotropies on small angular scales in cosmic
defect models \cite{recent}.  Ultimately,  measurements
should conclusively distinguish inflationary perturbations
from others \cite{test}.  Current evidence seems to favor
inflation \cite{recent}.

In this chapter, we will give an overview of some recent ideas
for implementing inflationary models in the context of supersymmetry.
Most of them are based on ``hybrid" \cite{hybrid,cope} inflationary models,
although there have been a few recent suggestions which try
to implement slow-roll in the context of a single inflaton field.
We will first review the motivation behind multifield inflation models,
and discuss some examples. We will then briefly discuss  recently
suggested single field models.

\section{Hybrid Inflation and Supersymmetry}
Before introducing supersymmetry into our discussion, let
us first consider the potential advantage to multifield inflation models.
To do so, we consider a toy model \footnote{I will
generally use the GRS \cite{rsg} conventions
for the slow-rolling inflaton field $\psi$ and
the  field which controls the energy density will be
denoted $\phi$.
The reader should be aware of other conventions existing.} with potential
\begin{equation}
V=(\phi^2-M^2)^2+\lambda \phi^2 \psi^2+m^2 \psi^2
\end{equation}
Notice that the $\psi$ potential is minimized when $\psi=0$,
at which point the $\phi$ potential is minimized with $\phi=M$,
where the potential energy $V=0$. On the other hand, it is
unlikely that $\psi$ starts at its vacuum value. In fact,
when the temperature exceeds the $\psi$ mass, the potential
is negligible and $\psi$ evolves extremely slowly towards its minimum.
Therefore in the early universe, one can reasonably expect large
values of $\psi$. If $\psi$ is greater than $\psi_c=\sqrt{2}M/\sqrt{\lambda}$,
and $\phi$ is sufficiently small, 
$\phi$ will rapidly move towards  the origin
where it will sit leading to a   vacuum energy of 
approximately $V=M^4$ (assuming this dominates the $\psi$ contribution
to the energy).
This  nonzero vacuum energy permits an inflationary stage.

In this model, inflation will end at around the time when the $\phi$
potential turns over, when $\psi=\psi_c$. In other implementations
of ``hybrid'' inflation \cite{linde}, it could be that 
inflation ends when the $\psi$ potential ceases to correspond to a slow-roll
situation. We will see an example of this shortly.

The above toy model is fine as a model of inflation.
In fact, multifield models seem to resolve very nicely
one of the major problems with the standard slow-roll
potentials; how can the potential be very flat
and then give rise to a rapid end and reheat? Having
two fields to control inflation solves this problem
beautifully. One field controls the vacuum energy,
whereas the other field essentially acts as a ``switch''
for inflation. 

 However, there 
are several obvious questions. First, why should the mass parameters
be small? And what sets the mass scales in the first place?
Nonsupersymmetric field theory cannot in general address this question.
Only for a Goldstone boson is there a reason to believe the mass
of a scalar is small; in general we would not expect a slow-roll
potential for the $\psi$ field.

Why can supersymmetry change this picture? First of all, flat
directions are natural in supersymmetric theories.  Not only
does supersymmetry protect against radiative corrections;
supersymmetric theories in general have a large moduli
space of flat directions which need not be put in by hand.
We should qualify what we mean by flat directions. In general,
we are referring to fields with no potential in the supersymmetric
limit, with no other fields away from their vacuum expectation value,
and with nonrenormalizable terms neglected. The presence of any
of these terms will in fact generate a potential, but one which
can in general be consistent with the requirements of inflation
if the curvature of the potential is sufficiently small (when
compared to the scale set by the vacuum energy).

The other interesting aspect of supersymmetric theories
is that in general they require at least one mass scale
which is distinct from the Planck scale. This is necessary
to account for the supersymmetry breaking scale, which
is lower than  the Planck scale if the standard low-energy
picture of supersymmetry breaking as accounting for stabilization
of the electrweak scale is correct.  The precise value of this scale
is model dependent. In hidden  sector models,
the new scale will be of order $10^{11} {\rm GeV}$, while
in models of supersymmetry breaking based on more direct
communication, the supersymmetry breaking scale  will be lower.
  The supersymmetry breaking scale, and in particular, the intermediate
scale,seems to be an obvious candidate for application
to inflationary models since it is associated with
nonvanishing vacuum energy density.  In some
supersymmetric models, other scales can appear.
Notable among these is the  Grand Unification (GUT) scale.  
Many models try to associate directly particle physics
models which incorporate a grand unified gauge theory
to inflation.

To account for density
perturbations, it is clear that there
needs to be some small number in the particle physics theory,
which could be a ratio of masses.  Much of the work on 
supersymmetric inflation has been focussed on trying to exploit
these mass scales to realize the necessary requirements of inflation.

It is useful to divide these efforts into two categories.
In one class of theories, the density fluctuations are roughly
of order $(M_G/M_p)^2$, whereas the second class of theories
only exploits the intermediate scale, and obtains density
fluctuations either as $(M_I/M_P)$ or as a result of various
parameters which might appear.  Here $M_G\approx 10^{16} {\rm GeV}$
is the GUT scale and $M_I \approx \sqrt{M_W M_{Pl}}$ is
the intermediate scale of order $10^{11} {\rm GeV}$
which determines the soft supersymmetry breaking
parameters in a hidden sector scenario for the communication
of supersymmetry breaking.  Notice
that the first class of models involves the scale $M_G$,
which is the VEV of some field and may or may
not set the magnitude of the potential energy density.
In the second case, $M_I$, we know it is associated
with a vacuum energy density. We will discuss each of these models
in turn.

Many of the models we discuss are given in the context
of global supersymmetry, although some models are incorporated into
a supergravity theory.  Before proceeding,
we mention two potential problems with supergravity inflaton models;
only some of the models presented below address these issues.
The first issue is that if $W$ is the superpotential and $K$ is
the Kahler potential,
the potential takes the form
\begin{equation}
e^{|K|}\left(\left(W_i+K_i W\right)K_{ij}^{-1}\left(\bar{W}_{\bar{j}}+K_{\bar{j}}
\bar{W}
 \right)-3 |W|^2\right) + D-{\rm terms}
\end{equation}
Since during inflation, the term in parentheses is nonzero if inflation
is due to nonvanishing $F$ terms, one will in general find
a large potential for any field appearing in the Kahler potential,
which will of course include the inflaton. This can destroy the
flatness of the inflaton potential and thereby destroy inflation
\cite{dfn,chrr,drt,bfs,cope}.

The other potential problem is that in the supergravity theory where
the cosmological contant at the desired minimum is cancelled
by a constant, one generically finds a deeper minimum
out at values of the field larger than $M_{Pl}$\footnote{I thank
Paul Langacker for stressing this problem.}. However,
if the superpotential is purely cubic in the fields this problem will not 
arise\footnote{I thank Gia Dvali for this comment.}. In general though,
it is difficult to know how to take this problem, as one is
generally treating the potential as a Taylor expansion, and
at field values beyond $M_p$ the theory is presumably no longer
valid. Furthermore, it is not clear that our world is in the global
minimum of the potential.

\section{Hybrid Inflation and High Scale Models}

The idea of hybrid inflation in supersymmetric theories
was studied in a seminal paper by Liddle, Lyth, Stewart, and Wands\cite{cope}.
Dvali,  Shaeffer, and Shafi [DSS] \cite{dss} pointed out the importance of considering quantum
supersymmetry breaking effects during inflation\footnote{Classical
supersymmetry breaking effects during inflation had been pointed
out in Ref. \cite{cope} and \cite{ewan1}.}, which they
then exploited to introduce an interesting model of inflation.
However, their model still required arbitrary mass scales.
 There were subsequent models in which the authors
tried to identify 
an appropriate mass scale. None of these models
are perfect, but might nonetheless have the germ of truth.

We first discuss the DSS model.  They have the superpotential
\begin{equation}
W=\kappa S \bar{\phi}\phi-\mu^2 S
\end{equation}
where $S$ is a singlet and $\phi$ and $\bar{\phi}$
transform under a GUT group.
Notice that when $\phi$ and $\bar{\phi}$ vanish, the $S$ field
is a flat direction.  The model also contains
an $R$-symmetry under which $S$ transforms which forbids
an $S^3$ term in the superpotential.  This is important
as it is essential that $S$ is a flat direction. 

Let us now consider the potential for this model. We have
\begin{equation}
V(S, \phi, \bar{\phi})=\kappa^2 |S|^2 (|\bar{\phi}|^2+|\bar{\phi}|^2)
+|\kappa \bar{\phi} \phi-\mu^2|^2+D-terms
\end{equation}
where the $D$-terms depend on the gauge representation
of  $\phi$ and $\bar{\phi}$.

Now at the supersymmetry preserving minimum, the $D$-term requirement
imposes $\phi=\bar{\phi}$, whereas the superpotential imposes
$\phi=\bar{\phi}=\mu/\sqrt{\kappa}$ and $S=0$.  However, 
in the early universe it is very likely that not all fields
were at their supersymmetry-preserving minimum.  In fact,
$S$ might have started off at a value $S>S_c=\mu/\sqrt{\kappa}$,
in which case the $\phi$ potential is minimized at vanishing $\phi$ 
and $\bar{\phi}$, where $V=\mu^4$.  In other words, this
is looking precisely like a hybrid inflation model, where
$S$ plays the role of $\psi$ and $\phi$ and $\bar{\phi}$
play the role of $\phi$ in our toy model.

Now naively it looks like $S$ is exactly flat, which would
be bad, since there would be no potential driving $\phi$
and inflation would never end.   However, this neglects
the fact that supersymmetry is broken during inflation! Here
the nonzero breaking is due to the nonvanishing $F$ term;
however we know this is generally true since inflation relies
on nonvanishing vacuum energy. In fact, in general this can 
be a problem in models with more than one mass scale. Since supersymmetry
is broken during inflation, this can be  unnatural as in a nonsupersymmetric
model. However, in this model, the quantum corrections 
introduce a potential for 
the $S$ field which is desired.

The consequence of the nonvanishing $F$ term and the breaking
of supersymmetry is that a potential for the $S$ field will be
generated through radiative corrections.  The one-loop effective
potential for $S$ is
\begin{equation}
\Delta V(S)=\Sigma {(-1)^F \over 64 \pi^2} M_i(S)^4 \log \left({M_i(S) \over 
\Lambda} \right)^2
\end{equation}
Here $M_i(S)$ are the $S$-dependent masses of the fields.
This effective interaction introduces a slope to the $S$ potential.
In fact, in this type of model, inflation generally ends when
slow-roll ceases to apply, rather than when the ``$\phi$" potential
turns over.

Let us consider this in more detail. Because of the supersymmetry
breaking $F_S$, the $\phi$ and $\bar{\phi}$ spectrum
do not respect supersymmetry. The scalars have
mass $\kappa^2 S^2 \pm \kappa \mu^2$, whereas the
fermion has mass $\kappa S$.  Substituting these $S$-dependent
masses into the effective potential, one derives the $S$ potential
at one-loop to be
\begin{equation}
V_{eff}(S)=\mu^4+{\kappa^2 \mu^4 \over 32 \pi^2}\left(\log{\kappa^2 S^2 \over
\Lambda^2} + {3 \over 2} \right)
\end{equation}

This model succeeds as a hybrid inflation model. However,
there are some important open questions. First, what is
the origin of the scale $\mu$.  To get the correct magnitude
of density fluctuations, it turns out that $\mu$ is of order 
the GUT scale. This means one might want to tie $\phi$ 
to the field with GUT mass and VEV. However, if we take $\phi$
to be an adjoint, there is too much global symmetry and
one obtains too many Goldstone bosons. Interactions which
violate this symmetry can also destroy inflation.  Alternatively,
one can take a GUT group like SU(6) and let $\phi$ and $\bar{\phi}$
be Higgs fields in the 6 and $\bar{6}$.
However, the VEV of this field is likely to be too low for a successful
GUT model.  So in summary, although the fact that the scale $\mu$
is of order the GUT scale is intriguing, in this basic model
it is tricky to realize the connection.

Another potential problem when the scale $\mu$ is high is
that the reheat temperature is likely to be too high, and
can cause problems with the gravitino constraint \cite{moroi}.
This is readily seen by a simple estimate assuming instantaneous
reheat.  Reheat occurs when the Hubble parameter $H$ is of order
of the inflaton
 width $\Gamma$. Since $H^2 \sim \rho/M_p^2 \sim T_R^4/M_p^2 \sim 
\Gamma^2$, we find $T_R \sim \sqrt{\Gamma M_p}$.  The bound
on the reheat temperature depends on the mass of the gravitino,
but is generally of order $10^{10} {\rm GeV}$.  If the inflaton
decays perturbatively, one expects $\Gamma \sim {\alpha \over 4 \pi} M_{inf}$.
If $M_{inf} \sim M_G$, this is clearly too big.  Even if the reheat
occurs through higher dimension Planck-suppressed operators, the reheat
temperature is probably too high, since it is of order $\sqrt{M_G^3/M_P}$.
This is not necessarily an insuperable problem, but it generally
requires a more complicated model. In the context of reheat,
it should be mentioned that there is still debate
over the role of parametric resonance in the decay of the inflaton; however this would generally only increase
the reheat temperature.  A further point is that the reheat bound assumes only gravitino couplings suppressed
by $M_{Pl}$. If the inflaton decays to particles in the sector
in which supersymmetry is broken, the rate for gravitino
production can be even larger and the reheat bound even stricter.
One can also  estimate a reheat bound
in gauge-mediated models of supersymmetry breaking \cite{murayama}.
One generally finds even more stringent bounds in this case, since
the gravitino is more strongly coupled.

A third problem is that our discussion so far has been in the
context of global supersymmetry.
 Planck-suppressed operators however cannot be neglected,
since the Hubble parameter itself is Planck-suppressed.  Without
some tuning, supergravity corrections can invalidate the conditions
for slow-roll. 

Subsequent models have tried to address the first type of problem,
namely the origin of the scale $\mu$ and some
have also addressed the third problem.   Various
authors \cite{dterm} (see also \cite{dterm1})
suggested $D$-term inflation. The idea is to generate the
scale ``$\mu$'' though a Fayet-Iliopoulos $D$-term.  They
envision a model with an anomalous field content in the low-energy
theory, with $n_+$ fields of charge 1 and $n_-$ fields
of charge -1.  The model contains a superpotential
\begin{equation}
W=\lambda_A X \phi_+^A \phi_-^A
\end{equation}
The potential for this model is then
\begin{equation}
V=\lambda^2 |X|^2 \left( |\phi_-|^2+|\phi_+|^2 \right)+\lambda_A^2 |\phi_+ 
\phi_- |^2 +{g^2 \over 2} \left( |\phi_+^i|^2+ |\phi_+^A |^2
-|\phi_-^A|^2 +\xi \right)^2
\end{equation}
If one looks at the potential along $\phi_+=0$, one
sees that this potential takes precisely the form
required for a successful hybrid inflation model.  
Here, $X$ plays the role of the $\psi$ field, and $\phi_-$
(or some linear combination) the  role of the $\phi$ field. 
One can work out the requirements for sufficiently long slow-roll
and for sufficient density fluctuations.

In this model, the vacuum energy density during inflation
is given by
\begin{equation}
V={g^2 \over 2} \xi^2
\end{equation}
This model has the nice
feature that because it is $D$-term inflation, one can control
supergravity corrections which would destroy slow-roll.  Recall
that there is no symmetry to prevent the quadratic terms in the
Kahler potential, which, in the presence of a nonzero $F$-term, will generate
an inflaton mass. The situation is better with $D$-term inflation;
however Lyth \cite{lythdterm} has pointed out that
even in this case, there can be large corrections is there
is no symmetry preventing quadratic corrections
to the holomorphic function of fields appearing in the gauge kinetic term.

One problem with $D$-term inflation is that it is difficult to get the
scale right. If
the parameter $\xi$ arises due to the Green-Schwartz
mechanism, it is about $g^2 {\rm Tr}Q M_{Pl}^2/192 \pi^2$,
and is  probably too big for the scale
set by density fluctations. One would need a small parameter
 to get the correct mass scale. 

Some authors have tried to address the question of getting
the correct size of the  $D$-term.  One possible
solution is that the $D$-term is generated at a scale
below the Planck scale. However, it is difficult to
see how this can be done without large $F$-term contributions
to the energy as well, destroying the initial
motivation for these models.

Matsuda \cite{matsuda}
pointed out that the strength of the gauge coupling determines
the magnitude of the $D$-term, and if this coupling is
dynamically determined, the $D$-term at the time of inflation
might be of a different size. He shows various ansatzes for the
dependence of the coupling on mass scale; unfortunately however
these are not motivated by any underlying physics. 
However, a realistic model of the scale dependence of the coupling
would require a solution to the problem of dilaton stabilization.  

March-Russell \cite{jmr} suggests that in a model in which
the string scale is reconciled with the GUT scale, one
could obtain better numbers for the size of the $D$-term.

Lyth and Riotto \cite{lythr} also tried to address the discrepancy
of scales required for $D$-term inflation. They point out that
the normalization of the magnitude of the $D$-term which
is required to agree with density perturbations  depends on
the slope of the potential at the end of inflation. In some
cases, this slope is given by the one-loop effective potential,
so by altering the number of fields coupled to the inflaton,
the slope can be increased. However adjusting the slope
by this (or any other mechanism) will only buy you at most
an order of magnitude (once consistency with the observations
on the spectral index $n$ are imposed) if one does not take
the coupling $g$ to be small.

Another interesting attempt to tie the $\mu$ scale
to a physical scale (here  a GUT scale) in the problem was made by
Dimopoulos, Dvali, and Rattazzi  \cite{ddr}.   Their model is based
on a quantum corrected moduli space, where the strong interaction
scale $\Lambda$ provides the scale for the overall energy density.
The model they give has a gauged SU(2) group with four flavors,
so there is a quantum modified moduli space.  The superpotential,
including the constraint, is
\begin{equation}
W_{eff}=A(Det M-\bar{B} B-\Lambda^4)+S({\rm Tr} M+{g' \over 2} {\rm Tr }\Sigma^2)
+{h \over 3} {\rm Tr} \Sigma^3
\end{equation}
where the last terms arise due to a tree-level superpotential,
and $Q \bar{Q}$ has been replaced by the confined meson
field $M$.  
 If the field $\Sigma$ is the adjoint field which breaks
$SU(5)$ down to the standard model, the scale $\Lambda$
should be of order the GUT scale, which works well
for producing density fluctuations.  

Inflation does not involve the $\Sigma$ field until the end. Initially,
the model works as with other hybrid inflation models. $S$ is a flat direction.
When $S$ is big, there is a mass, and $M$ sits at zero, so there is
nonzero vacuum energy. In this model, inflation ends at a nonzero value
of $S$ and $\Sigma$ and SU(5) is broken. In principle, inflation could also
end with  $\Sigma$ zero although the authors of Ref. \cite{ddr}
argue that this is not the case.  

There are other models which try to incorporate hybrid
inflation into a GUT model. For example, Covi, Mangano, Masiero,
and Miele \cite{cmmm} implement hybrid inflation in
 an SU(5) model with an additional singlet, and an arbitrary
parameter $\mu$ which they need to take of order the GUT scale.
Another model is given by Lazarides, Panagiotakopoulos,
and Vlachos \cite{lpv}, who use a nonrenormalizable potential
to introduce a slope to the inflaton field (which was
given by supersymmetry breaking parameters in other models).

One potential worry with any model based on SU(5) is that most
such models do not solve the doublet-triplet splitting problem,
and are therefore unrealistic unless a severe fine-tuning is imposed.
Since the point of inflationary model building is to eliminate small
parameters, this is a less than satisfactory situation.

An even  more severe problem in models which really
try to tie inflation to   an SU(5) GUT of the real world 
was pointed out by Dvali, Krauss, and Liu \cite{krauss}. They point
out that in SU(5) models with an adjoint which gets a nonzero VEV,
there will be two choices of vacua, one in which $SU(4) \times U(1)$
is preserved, and one in which $SU(3) \times SU(2)$ is preserved
They parameterize the vacua with three parameters, the overall
scale of the symmetry breaking, and two angles (or orbit parameters).
When the inflaton field  (here we mean the field
whose potential generates the vacuum energy density)
begins to evolve away from its inflationary
value, only one potential minimum is present, namely that
corresponding to the bad $SU(4)\times U(1)$ vacuum.
They argue further that the field will never make its way
to the desired $SU(3)\times SU(2)$ vacuum.  Furthermore,
whatever vacuum is chosen, the transition happens {\it after}
inflation, so the monopole problem is not solved. So without
embellishment, the simplest hybrid inflationary models
based on SU(5) GUTS are not successful. These authors
suggest possible resolutions which involve somewhat more
complicated theories. It is also possible that in a model
such as that of Ref. \cite{ddr} that a noncanonical Kahler
potential invalidates the energy argument and that the
suitable vacuum is obtained.

Most authors do not address the question of reheat.
Lazarides \cite{lss} suggests  a decay to a second generation neutrino
to avoid too big renormalizable couplings. It is hard
to think of a natural decay mode without small coupling
which can avoid a high reheat temperature and
overproduction of gravitinos. An alternative
proposal of Dimopoulos and Dvali is that
reheat is delayed by rolling along a flat direction
\cite{dvaliprivate}; however it is necessary to ensure
that no other dangerous perturbations will be produced.
 It could be that there
is some late entropy release which invalidates the gravitino
bound; this might be required to solve the Polonyi problem
\cite{polonyi} in any case. It is clear that the question of the
high reheat temperature should be addressed in these high 
scale models.
 
So to summarize, models of hybrid inflation based on a high
$H$ scale seem close to working. $D$-term inflation doesn't quite
get the scale right, but is close. Models based on SU(5)
generically suffer from the problem outlined in \cite{krauss}
which is unfortunate since it makes the very nice coincidence
of scales less useful. However, it is not impossible to make
these models work, and further advances might be forthcoming.
 
\section{Low-Scale Models}
We now go on to discuss another very promising
 class of models, which do
not introduce a high scale Hubble parameter, but try to implement
successful inflation only by assuming the existence of soft-supersymmetry
breaking in a hidden sector. 
These models are intended as illustrations of how the moduli
fields can be employed in a hybrid inflation scenario
without strong or unnatural assumptions on the particle physics model.
With the specific cases that were discussed, one can
identify  distinguishing characteristics of this
class of model, which should be testable.
 
The goal of  Randall, Solja\v c\'ic, and Guth (RSG) was
to construct models of inflation using {\it only} the intermediate
mass scale $M_I$ and the Planck scale $M_p$. The aim
was to see whether a natural model could be simply constructed
which employed moduli fields with soft supersymmetry breaking
masses. The essential observation is that the temperature
fluctuations observed by COBE do not necessarily require
high scale inflation, since the formula for density fluctuations
actually depends both on the magnitude of the potential and its
slope.   Taking the potential  quadratic at the time
density fluctuations relevant to physical scales are formed, the
formula for density fluctuations is 
\begin{equation}
{\delta \rho \over \rho}\approx {H^2 \over \dot{\psi}}\approx{H^3 \over
m^2 \psi}
\end{equation}
where $\psi$ is the slow rolling field. Now
if the energy density during inflation is $M^4$, the Hubble
parameter is of order $M^2/M_p$. On the other hand,
if the $\psi$ mass arises from hidden sector supersymmetry
breaking, it is $m \approx M^2_I/M_{Pl}$.   The assumption in this
class of models is that $M \sim M_I$, in which case
\begin{equation}
{\delta \rho \over \rho} \approx {H \over \psi}
\end{equation}
{}From this equation, it is clear that the magnitude of density
fluctuations depends on the value of $\psi$ when inflation
ends, which for hybrid inflation  models is essentially $\psi_c$.

In Ref. \cite{rsg}, two types of potentials were considered.
The first class of models  assumed the fields
were coupled through a higher dimension operator derived
from the superpotential
\begin{equation}
W={\phi^2 \psi^2 \over 2 M'}
\end{equation}
where $M'$ is a relevant physical mass scale.  Perhaps the most
natural possibility is $M_p$. However, it is conceivable there
are heavy particle exchanges so that $M'$ can be identified
with the GUT scale $M_G$ or $M_I$.  

The other class of model assumed there was a potential coupling
the two fields involved in hybrid inflation of the form
\begin{equation}
V={\lambda \over 4} \psi^2 \phi^2
\end{equation}
Such a coupling could arise for example if the superpotential
coupled together three fields $W=\chi \phi \psi$,
where $\chi=0$ during inflation.  This is in fact something
which happens quite naturally, even in the context of the MSSM.
For example, if $\psi=\bar{u} \bar{d} \bar{d}$ and $\phi=H_u H_d$,
$W=\lambda_u Q_u H_u \bar{u}$, one realizes this situation.
In fact, nonstandard GUT models \cite{rsg,mygut} can realize
this potential in a way consistent with the inflationary constraints.
In this model, the magnitude of density fluctuations will be
set by a Yukawa coupling; it is important to recognize that
this might well be significantly less than unity.

The complete  specification of the model requires
the potential for the $\phi$ and $\psi$ fields (apart from their
mutual coupling).  Note that these potentials arise due to
soft supersymmetry breaking and therefore  should be  characterized
by   potentials of the form $M_I^4 g(\phi/{M_p})$ where $g$ is a function
with a Taylor expansion with coefficients of order unity. To
realize the hybrid inflationary scenario,  the potential
for $\phi$ is taken as $V=M^4 \cos^2(\phi/\sqrt{2} f)$ and the potential
for $\psi$ is taken as ${1 \over 2 } m_{\psi}^2 \psi^2$. To be consistent
with the  requirement that these are moduli
fields with a potential generated by soft supersymmetry
breaking, we would want to find $M\sim M_I$
and $m_{\psi} \sim M_I^2/M_p$.  There has been much
confusion over the very specific-looking form taken for the $\phi$
potential.  Indeed, all that is relevant to inflation are the first
two terms in the Taylor expansion! Only when inflation ends and $\phi$
moves from zero are the other terms relevant.  This is simply
a compact way of writing a function which has the correct negative
curvature at the origin and zero vacuum energy for the true vacuum.
It is interesting however that exactly such a potential could be produced
by a nontrivially coupled pseudo-Goldstone mode \footnote{I thank Gia
Dvali for sharing this observation of Lawrence Krauss and himself.}.
However, this precise form of the potential is not at all essential,
so the field $\phi$ can be any moduli field with negative mass squared
and a supersymmetry-breaking source for its potential.

This model realizes very nicely the hybrid inflation scenario.
Depending on the form of the soft-supersymmetry breaking
potential and the couplings between moduli fields, it is
very likely one can find suitable candidates for inflation.
The major distinguishing characteristic of this type
of model is that the $\phi$ field is light. Therefore,
the dynamics controlling the end of inflation is very different.
In many other models, the $\phi$ mass is large, so it
very quickly rolls to its true minimum once
inflation has stopped. In the RSG models,
the field $\phi$ spends more time moving
primarily  due to de Sitter fluctuations,
subsequent to which it rolls classically. Because
of the initial motion when the
field is moving relatively slowly, there is   a spike
in the density fluctuation spectrum.  This spike can be
interesting (or dangerous) in that it will lead to more structure
on small scales. However, it is too large to be present in observed
density fluctuations on scales from about $1$ Mpc to $10^4$ Mpc,
which gives a constraint on how quickly inflation must end. Including
this constraint as well as the constraint from density fluctuations,
one finds that the parameters of this model are such that
it works rather well, with mild tuning depending on the particular
model (numbers as small as 0.01 might be required; see \cite{rsg} for
details). That is, the original goal, to motivate
the parameters by supersymmetry breaking scales, can be
reasonably well accomodated.

It should be noted that the derivation of the
parameters of the spike in the density perturbation
spectrum is subtle.  In Ref. \cite{rsg},
the calculation was based on using the Fokker-Planck equation
to establish the $\phi$ mean $\phi$ distribution
and the time delay given by a fluctuation in the $\phi$
field.  Garcia-Bellido, Linde, and Wands\cite{gblw}
objected to the calculational method of \cite{rsg} and
instead calculated the fluctuations associated with
each element of the ensemble assuming it was classical. However,
it can be shown \cite{grsu} that the method used is not valid,
though the original \cite{rsg} calculation needed to be
improved to account for deviation from slow-roll and
for a more exact calculation of the fluctuations for a massive field.

Stewart \cite{qf1,qf2} has also constructed low-scale  models for which
the small tuning required to get a sufficiently flat
potential is not required and for which the spike 
will have different properties.  He points out that
once quantum corrections are incorporated, there can
be a special point (or more than one) where the potential
is particularly flat. Initial
conditions are probably different for this class of
model; one relies on entering a phase of eternal
inflation from which one will enter the desired
hybrid inflationary phase.

In summary, the low-scale inflation models can have quite distinctive
features, and do not have the problems associated
with introducing a high scale in a particle physics context.  
However,  they might involve a small amount of fine-tuning;
on the other hand they might also involve a small parameter which is
present.  The spike is generically
a test of the models; however if the field is rolling
quickly at the phase transition, as is true for Ref. \cite{qf2},
this might be lessened; futher work is needed for these
models to establish the detailed form of the spike.
In general, the low-scale models are well motivated
and worthy of further investigation.

Before discussing further models, it is worth noting a distinguishing
feature of many hybrid inflation models, those
with a mass for the inflaton (like the GRS models),
 namely the fact
that the index $n$ is generally bigger than 1. Furthermore,
by measuring the ratio of tensor to scalar perturbations, 
one can in principle (because
it is a difficult measurement) distinguish high and low scale models.
The deviation of the index $n$ from 1 measures the scale
dependence of density fluctuatons. It can be determined
from the potential at the time the relevant perturbation
leaves the horizon from the formula
\begin{equation}
n=1-3 \left({V' \over V} \right)^2+2 {V'' \over V}
\end{equation}
whereas $R$, the ratio of tensor to scalar perturbations,
is given by
\begin{equation}
R \approx 6 \left ( {V' \over V} \right)^2
\end{equation}
where in these  equations we have set $M_p$ to unity.
One can obtain  interesting qualitative information from these
formulae.  First consider the quantity $R$.  If we can approximate
$V'$ by a mass term near the end of inflaton, we have
$V'/V\sim m^2 \psi M_p/H^2 M_p^2$. So if $H \sim m$, which
is often the case, we find that $R$ is negligible unless
$\psi \sim M_p$. As we have argued, models with low $H$
can achieve adequate density perturbations if $\psi$ is
small (much less than $M_p$) at the end of inflation. We conclude
that these type of models will always have neglibile $R$.
For other models, with $\psi$ closer to $M_p$, it
is a detailed question whether $R$ can be measurable.

Notice also that when $V'/V$ is neglibible, the sign of the
mass squared term at the end of inflation determines whether
$n$ is bigger, or less than unity.  So models for which the
inflaton field  rolling towards the origin will have $n$ bigger
than 1. The RSG models are of this type, as are  hybrid
inflation models with a mass term determining the evolution
of the inflaton.

It should be noted that there are many models where the potential for
the inflaton is not determined by a mass term.
An interesting example of hybrid inflation 
which he dubbed ``Mutated Hybrid Inflation" for which the index $n$
is less than 1 was given by Stewart \cite{mhi}.  He considers
a toy model where inflation occurs along a nontrivial trajectory
in field space. The net result is that along this trajectory,
the potential can be written as a polynomial function of the
inverse field, and the index $n$ can be shown to be less than 1.
Generalizations of this idea and other mechanisms for producing
an index less than 1, again in toy models, was given in Ref. \cite{lythstew}.

\section{Single Field Models}

Aside from the many models of hybrid inflation based on supersymmetry,
there are a couple of single field inflationary models worthy of note.
By single field, we do not mean there is only one field in the potential,
but that the inflationary dynamics can be viewed in terms of a single
field (as opposed to hybrid inflation models).
Garcia-Bellido \cite{bellido} observed that the potential
given by an $N=2$ SU(2)  gauge theory with supersymmetry breaking 
\cite{lag} 
 incorporated takes a form which looks  remarkably like a slow-roll
potential along a particular trajectory in field space.  However,
the scale of supersymmetry breaking which is required has no particle
physics motivation, so it is not yet clear if this can be tied
to a particle physics theory of our world.

Another  single 
field model is that of Adams, Ross, and Sarkar \cite{ars}.   
They are interested  in the problem
of the large quadratic terms which can be present in supergravity theories. They argue  that
there can be special points where the quadratic terms vanish,
and these can be quasi-fixed points in the evolution of the field.

Another  paper which addresses the issue of large supergravity
corrections is by Gaillard, Murayama, and Olive \cite{gmo}.
They observe that at tree-level, the mass term for the inflaton
which occurs generically in supergravity theories
is absent if there is a Heisenberg symmetry.  Although
there is no symmetry reason, it is claimed that gravitational
interactions preserve the symmetry (based
on a one-loop calculation), so that the potential
can be calculated from gauge and superpotential interactions.
A more complete realization of this scenario could be interesting.

Stewart\cite{ewan1} also addressed the issue
of large supergravity corrections to the inflaton mass.  He
identified  conditions, which when imposed on the superpotential,
guarantee the absence of such corrections. These conditions
are $W= W_\psi=\phi=0$ (in the GRS naming convention) during inflation,
as well as some conditions on the Kahler potential.
He argues that such potentials might arise naturally in superstring theories.

In fact,  Copeland, Liddle, Lyth,  Stewart, and Wands\cite{cope} had
initially pointed out that supergravity corrections to the mass can
cancel if there is a minimal Kahler term.  Linde and Riotto \cite{lir}
make the assumption that nonminimal terms which  would
destroy this cancellation are small, and  then consider the model
with both one-loop and gravitational effects taken into account.

In summary, there are currently many ideas on how to use
the naturalness property of supersymmetry to provide candidates
for inflaton fields. There are some clever ideas involved
in high scale models; however the $D$-term models generally
give too large density fluctuations while the GUT models
often lead to the wrong vacuum following inflation.  These models
might however be incorporated into more complete and realistic
models in the future.  The low scale inflation models have the
advantage that they require no new mass scales aside
from that which was already required to give supersymmetry
breaking parameters of order the weak scale in a hidden sector
scenario.  They provide an interesting signature of a spike in the
spectrum so they should be subject to experimental verification in
the future.  Other objections might  include
the fact that this inflation is relatively late,  so
this might mean some prior nonstandard evolution (like
a previous inflationary phase)  is required.
 These models require mild tunings; presumably even this
is not necessary  if one will compromise with more complicated scenarios.  It is intriguing that new models of particle physics might also
lead to new potentials  that could provide slow-roll.
Exact superpotentials in the strong interaction regime
often take nonpolynomial forms which would not have been
anticipated on the basis of weakly coupled renormalizable
field theories.  Other models which fall outside the
range of the supersymmetry-motivated field theories
we have considered here include dilaton-based
inflation \cite{dilaton} and  string-theory-motivated-domain walls   
as the seed for inflation \cite{banksco}.

Although there are many ideas, 
  it should be remembered that there
are many requirements for a good inflationary model. Given
the indirectness of many cosmological constraints, it
is remarkable how constrained models are.
 Requirements include  a sufficiently long period of inflation,
a mechanism for ending inflation sufficiently quickly to
reheat to temperatures higher than the weak scale (this
might be too stringent but in alternative scenarios one
needs a mechanism for baryogenesis), a reheat temperature
sufficiently low not to overproduce gravitinos, consistency
with a spectral index which does not deviate by more than 20\%
from 1 (the exact constraint is subject to interpretation), and hopefully
no ad hoc scales or small numbers. There are  only a 
few models which   meet all these criteria.  And we have not even
addressed the many issues of how inflation fits
into a more complete picture of the cosmology of the
early universe which includes baryogenesis \cite{adb} and a
solution to the Polonyi problem \cite{polonyi}.  So
 despite the many recent advances, the field
still remains fertile, and it would not be surprising to
see new and more compelling models of inflation in the future.

\noindent{\bf Acknowledgements} I am very grateful to  Ewan Stewart and Chris Kolda
for discussions and  their comments on the manuscript. I also thank
Gia Dvali, Marc Kamionkowski, John March-Russell, and
Riccardo Rattazzi for their comments. I also thank Princeton University
and the Institute for Advanced Study for their hospitality while
this work was completed.  This work is supported in part by funds provided
by the U.S. Department of Energy (D.O.E.) uncer cooperative agreement
 \#DF-FC02-94ER40818.

\end{document}